\documentclass[aps,twocolumn,showpacs]{revtex4}
\usepackage{epsfig}
\newcommand{\be}{\begin{equation}}
\newcommand{\ee}{\end{equation}}

\begin{document}

\title{Critical Percolation in High Dimensions}

\author{Peter Grassberger}

\affiliation{John-von-Neumann Institute for Computing, Forschungszentrum J\"ulich,
D-52425 J\"ulich, Germany}

\date{\today}

\begin{abstract}
We present Monte Carlo estimates for site and bond percolation thresholds 
in simple hypercubic lattices with 4 to 13 dimensions. For $d<6$ they are 
preliminary, for $d\ge 6$ they are between $20$
to $10^4$ times more precise than the best previous estimates. This was achieved 
by three ingredients: (i) simple and fast hashing which allowed us to simulate
clusters of millions of sites on computers with less than 500 MB memory; 
(ii) a histogram method which allowed us to obtain information for several $p$
values from a single simulation; and (iii) a variance reduction technique
which is especially efficient at high dimensions where it reduces error bars
by a factor up to $\approx 30$ and more. Based on these data we propose a new 
scaling law for finite cluster size corrections.
\end{abstract}

\maketitle

In spite of decades of intensive studies \cite{sa94}, percolation remains
an active subject of research.
%
While there has been enormous progress in understanding percolation in 2 
dimensions \cite{cardy}, mainly because of conformal invariance, progress in 
high dimensions has been much slower. It has been known since long time that 
$d=6$ is the upper critical dimension \cite{sa94}, and expansions of $p_c$ in 
$1/(2d-1)$ have been given already more than 20 years ago. But up to now there 
exists no detailed numerical study of logarithmic corrections in $d=6$, finite
size corrections are not yet understood for $d>6$, and even
numerical estimates of $p_c$ in $\ge 6$ dimensions are very poor. One reason 
for this is obviously that straightforward simulations of large lattices in 
$\ge 6$ dimensions require huge amounts of fast memory. This lack of stimulus 
by numerical verification certainly was part of the reason for the slow
analytical progress.

It is the purpose of the present note to improve on this situation by presenting 
precise numerical estimates of $p_c$ (and of finite cluster size corrections) 
for site and bond percolation on simple hypercubic lattices with $d=6$ to $d=13$.

Our main results are summarized in Table 1, where we also include preliminary 
results for $d=4$ and $d=5$. We also give the best
previous estimates for $p_c$ and expansions in $1/(2d-1)$. We shall discuss 
them later in more detail, but here we just point out that our new estimates 
are vastly better than all previous ones. They were possible, with rather 
modest effort (we used only fast PCs and Alpha work stations, with 
altogether ca. $10^3$ CPU hours), due to several important ingredients:

{\bf 1)} We used as basic routine a standard breadth-first version of Leath's algorithm
which simulates single clusters. We do {\it not} use the popular Hoshen-Kopelman
method since that would require prohibitively large memory if we want to simulate
large clusters. In Leath's method, one writes the coordinates of each cluster site
(which consist of a single integer -- see item 2 below) into a first-in-first-out
queue Q, where each new entry represents a newly wetted neighbor of the oldest 
entry in the queue.

{\bf 2)} We used a simple but very efficient form of hashing \cite{s90} for storing 
the information whether a site has already been wetted or not. On Compaq Alpha work 
stations with 64 bit long integers, we labelled lattice sites by a single long 
integer. Using as lattice size $L$ an odd number slightly smaller than $2^{64/d}$, 
we label the neighbors of site $i$ as $i\pm 1, i\pm L,\ldots i\pm L^{d-1}$. If we 
want to simulate bond percolation clusters with roughly $N$ sites, we find first the 
power of 2 nearest to $N$, $2^k \approx N$, and use it to obtain for each site
$i$ its key $m_i=(i\; {\rm mod}\; 2^k)$ (notice that this is done most efficiently by 
bitwise AND). Assume now that site $i$ with key $m_i$ is the $n$-th site wetted.
Then an entry is written into the $m_i$-th element of an array of pointers S of size 
$2^k$.  This element points to the $n$-th
element of a structure (L,Q) where Q is the above queue and L is a linked list.
In Q, the coordinate $i$ is stored. The element of L remains empty, if the key $m_i$ 
had not been encountered before. Otherwise, if some other site $j$ with the same key 
$m_j=m_i$ had been wetted in an earlier step $n' < n$, the old element of S (which 
had pointed to $n'$) is written in the $n$-th element of L.
In this way we can deal with virtual lattices of $2^{64}$ sites, using $2^k + 2N_{\rm max}$
storage places, where $N_{\rm max}$ is an upper bound on the size of clusters to 
be simulated. The algorithm is slightly different for site percolation where 
a tested site has to be excluded from further growth even if it is not wetted, 
in contrast to bond percolation. It also has to be modified on machines with only 32 bit 
long integers where a pair of numbers replaces $i$ and a pair of co-prime odd numbers
$L_1$ and $L_2$, both slightly smaller than $2^{32/d}$, replaces $L$.

This is not as storage efficient as the recent algorithm of \cite{pzs01}.
But it works with usual (pseudo-)random {\it number} generators (we used the 4-tap 
generator with period  $2^{9689}-1$ of \cite{ziff}), while the algorithm of \cite{pzs01}
needs a random {\it function} generator. The most easily available random function 
generator today is the Data Encryption Standard \cite{press} which is rather slow when 
implemented in software and of unproven quality for this application (it was developed
for entirely different purposes, and lacks any published theoretical justification).

\begin{table*}
\begin{center}
\begin{tabular}{|r|c|c|c|c|c|c|c|} \hline
      &   \multicolumn{3}{c|}{bond}              &     \multicolumn{4}{c|}{site}           \\ \hline
      &              & \multicolumn{2}{c|}{previous}  &         & \multicolumn{3}{c|}{previous} \\ \cline{3-4} \cline{6-8}
  $d$ &   present    & best estimate & Eq.(\ref{seb}) & present & best estimate & Eq.(\ref{ses}) & Eq.(\ref{sesa}) \\ \hline
   4  & .1601314(13)   & .160130(3) \cite{pzs01}& .15666092 & .1968861(14) & .196889(3) \cite{pzs01}& .19304456 & .19880605\\
   5  & .118172(1)   & .118174(4) \cite{pzs01}& .11664888 & .1407966(15) & .14081(1) \cite{pzs01}& .13793629 & .14004471\\
   6  & .0942019(6)  & .09420(1) \cite{amah90} & .09365356 & .109017(2) & .1079(5) \cite{m98}  & .10754047 & .10848530\\
   7  & .0786752(3)  & .078685(3)\cite{amah90} & .07847711 & .0889511(9) & .08893(2) \cite{sz00}  & .08823220 & .08871655\\
   8  & .06770839(7) & .06770(5) \cite{amah90} &  .06763062  & .0752101(5)  &      --                   & .07485431 & .07512757\\
   9  & .05949601(5) & .05950(5) \cite{amah90} &  .05946233  & .0652095(3)  &      --                   & .06502556 & .06519119\\
  10  & .05309258(4) &    --       &  .05307663  & .0575930(1)  &      --                   & .05749265 & .05759880\\
  11  & .04794969(1) &    --       &  .04794152  & .05158971(8) &      --                   & .05153203 & .05160316\\
  12  & .04372386(1) &    --       &  .04371939  & .04673099(6) &      --                   & .04669616 & .04674559\\
  13  & .04018762(1) &    --       &  .04018504  & .04271508(8) &      --                   & .04269312 & .04272853\\ \hline

\end{tabular}
\caption{Estimates of $p_c$ for bond and site percolation in $d=4$ to $d=13$. Numbers in
round brackets are single standard deviations, square brackets refer to the citations at 
the end of the paper. For $d>9 $ the best previous estimates backed by theory were 
given by the (presumably asymptotic) expansions (\ref{seb}) and (\ref{ses}), while 
Eq.(\ref{sesa}) was a heuristic guess. The estimates for $d=4$ and $d=5$ are preliminary,
since we do not yet understand the important corrections to scaling in these cases (all 
error bars in this paper include plausible worst case estimates of systematic errors).}
\end{center}
\end{table*}

{\bf 3)} In order to estimate cluster statistics for several values of $p$ from a single
run at nominal value $p_0$, we use a trick similar to the histogram methods used by 
Dickman \cite{dickman} for the contact process (see also \cite{balle}). If a cluster 
with $n$ wetted sites and $b$ non-wetted boundary sites was generated with nominal value 
$p_0$, it contributes to the ensemble with $p_0$ replaced by $p$ with weight
\be
   W = (p/p_0)^n((1-p)/(1-p_0))^b .
\ee
Instead of collecting histograms for cluster numbers with fixed $n$ and $b$ (which 
would have led to excessively large arrays) we calculated on the fly three distributions:
One for the nominal $p_0$ (which was chosen close to $p_c$ as estimated from short test 
runs and from Eqs.(\ref{seb}) resp. (\ref{ses})) and two for neighbouring values $p_\pm 
= p_0 \pm \delta p$, using Eq.(1) for the latter. Observables at $p$-values in between 
(including $p_c$) were obtained by geometric (i.e. linear in logarithm) interpolation. 
Having three values of $p$ instead of just two allowed us to check that the error due 
to the interpolation was negligible.

{\bf 4)} Our main observable will be the number $M(t)$ of wetted sites with ``chemical 
distance" $t$ from the seed of the cluster (i.e. the number of sites infected at time 
$t$, if cluster growth is interpreted as spreading of an epidemic). For $d>6$ we 
expect its average $\langle M(t)\rangle$ to become a constant at the critical point, 
since the process is basically 
a branching process with small corrections. But instead of using $\langle M(t)\rangle$ 
itself, we obtain a less noisy signal by the following trick which would give the 
{\it exact} ensemble average of $M(t)$ if the cluster growth indeed were a branching 
process \cite{footnote}. 

Let us assume we have a (still growing) cluster $C$ with $M(t)$ sites wetted at step 
$t$, and denote by $M^+(t)$ the number 
of free neighbors, i.e. the number of sites which {\it can be} wetted at step $t+1$.
The actual number wetted will fluctuate, but the expected average number, conditioned 
on $C$ and thus also on $M(t)$, is exactly given by 
\be
   {\bf E}[M(t+1)|C] = p M^+(t).  
\ee
Thus the expected geometric increase of the number of wetted sites, still conditioned 
on $C$, is 
\be
   {\bf E}[M(t+1)/M(t)|C] = p M^+(t) / M(t)                                  \label{mhat1}
\ee
and its weighted sample average over all clusters is 
\be
   r(t) \equiv {\sum_C M(t) {\bf E}[M(t+1)/M(t)|C] \over \sum_C M(t)} = 
                   {p \langle M^+(t) \rangle \over \langle M(t)\rangle}.     \label{mhat2}
\ee
Our estimate for the true ensemble average of $M(t)$ is then finally
\be
   \widehat{M(t)} = \prod_{t'=0}^{t-1} r(t').                                 \label{mhat3}
\ee
Since we measured also the direct estimate $\langle M(t)\rangle$ and the 
(co-)variances of both estimates, we can also compute the variance of any 
linear combination of both. For $d=4$ and $d=5$, where both variances are 
comparable and the covariance is negative, a substantial achievement is obtained 
by taking as the final estimate the linear combination with the smallest variance.

An expression similar to Eq.(\ref{mhat3}) can be obtained also for the rms. 
radius, if we replace 
the ratios in Eqs.(\ref{mhat1},\ref{mhat2}) by differences and the product 
in Eq.(\ref{mhat3}) by a sum. This would also be exact and non-fluctuating 
if the cluster growth were a branching process with translation invariance.

The variance reduction due to Eq.(\ref{mhat3}) is largest for small $t$. 
Yet, for bond percolation in $d=11$, it gave even for the largest $t$ (=200) 
a factor $\approx 1/1000$ over using just $\langle M(t)\rangle$. 
For $d=6$ and $t=2000$, the reduction was still by a factor $\approx 140$. 
Indeed, there were substantial improvements even for $d=4$ and 5, while 
the improvement in $d=3$ was marginal. For site percolation the improvements 
were similar but somewhat less dramatic.

We should note that we calculated also $P(t)$, the probability that a cluster 
survives at least $t$ steps (i.e. has ``chemical radius" $\ge t$), the cluster 
size distribution $P(n)$, and the spatial extent of clusters with $n$ sites.
All of them gave vastly more noisy signals (since we could not use a similar 
variance reduction trick as for $M(t)$) and were not used in estimating 
the critical point.

Results for $\widehat{M(t)}$ are shown in Figs.1 to 3 for $d=6, 7$, and 11.
In all these  figures we show results for bond percolation. Results
for site percolation are similar albeit somewhat more noisy. In the first two 
cases we checked explicitly that no cluster was larger than the virtual 
lattice size $L$ (which was $>500$ in both cases), so there are strictly 
no finite lattice size effects. For $d\ge 9$ this was no longer possible for 
the cluster sizes used here (typically up to $10^4 - 10^6$ sites), but we can 
easily convince ourselves that also there finite size effects are negligible. 

\begin{figure}
\psfig{file=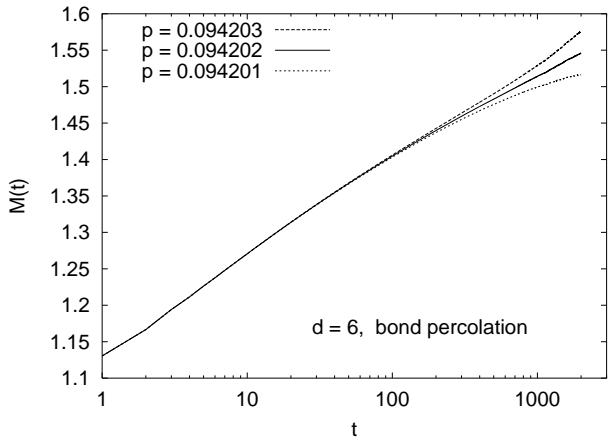,width=6.0cm,angle=270}
\caption{Plot of $\widehat{M(t)}$ versus $\ln t$ for bond percolation in $d=6$.
Statistical errors are smaller than the width of the curves. The main uncertainty
in pinning down $p_c$ comes from the non-obvious and somewhat subjective 
extrapolation to $t\to\infty$.}
\label{fig1.ps}
\end{figure}

In each of the 3 figures the critical point $p_c$ is characterized by 
$d\widehat{M(t)}/d t \to 0$ 
for $t\to\infty$. For $d>6$ we also have $\widehat{M(t)} \to const$ for $p=p_c$,
while we see a logarithmic divergence in $d=6$ as predicted by the renormalization
group \cite{egg78} (see Fig.1). Unfortunately, the detailed behaviour of $M(t)$
in $d=6$ has not yet been calculated, though the results of \cite{egg78} and 
the fact that $\nu_t$ (the exponent controlling the correlation time) is 1,
suggest $M(t) \sim [\log t]^{2/7}$ to leading order. Therefore, and since it is 
notoriously difficult to verify logarithmic terms 
(see, e.g., \cite{g97,ghs94,po01}), we have not attempted any detailed analysis.

\begin{figure}
\psfig{file=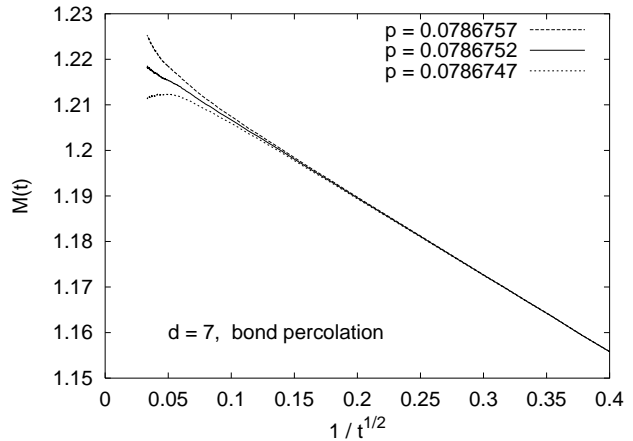,width=6.0cm,angle=270}
\caption{Plot of $\widehat{M(t)}$ versus $t^{-1/2}$ for bond percolation in $d=7$.
Statistical errors are always smaller than half the distances between neighboring
curves.}
\label{fig2.ps}
\end{figure}

\begin{figure}
\psfig{file=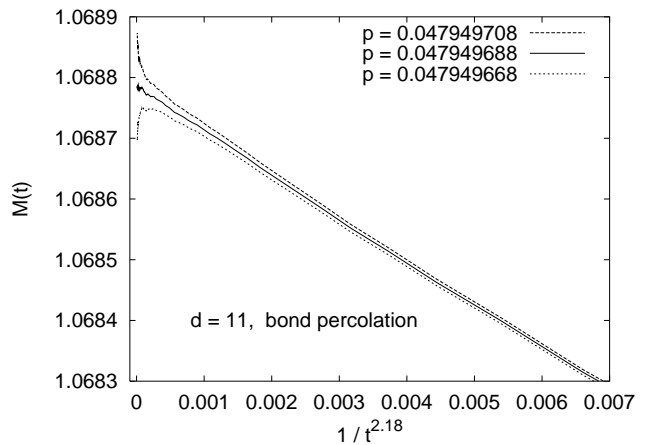,width=6.0cm,angle=270}
\caption{Plot of $\widehat{M(t)}$ versus $t^{-2.18}$ for bond percolation in $d=11$.
Statistical errors are again smaller than half the distances between neighboring
curves. The exponent 2.18 is chosen since it gives the straightest line.}
\label{fig3.ps}
\end{figure}

From Figs.2 and 3 we also see that corrections to scaling decrease strongly 
with dimension for $d>6$. In Fig.2 we see a straight line for $p=p_c$ when 
plotting $\widehat{M(t)}$ against $1/\sqrt{t}$, showing that the leading 
correction term is $\propto t^{-0.5}$ in $d=7$. Similarly, a straight line is
obtained for $d=11$ when using $t^{-2.18}$ (Fig.3). All these (and similar results 
for other values of $d>6$ and for site percolation, not shown here) strongly 
suggest anomalous scaling 
\begin{equation}
   M(t) = M_\infty - const / t^{\omega(d)},                 \label{mt}
\end{equation}
similar to the scaling for self avoiding walks in $d>d_c$ found in \cite{po01}.
But while the exponents were simply $(d-d_c)/2$ in \cite{po01}, they 
seem to depend less trivially on $d$ in the present case -- although we cannot 
exclude the possibility that $\omega(d)=(d-d_c)/2$ also here, and the observed 
deviations are due to higher order corrections. The latter is indeed suggested
by the results of \cite{aah84}.

\begin{figure}
\psfig{file=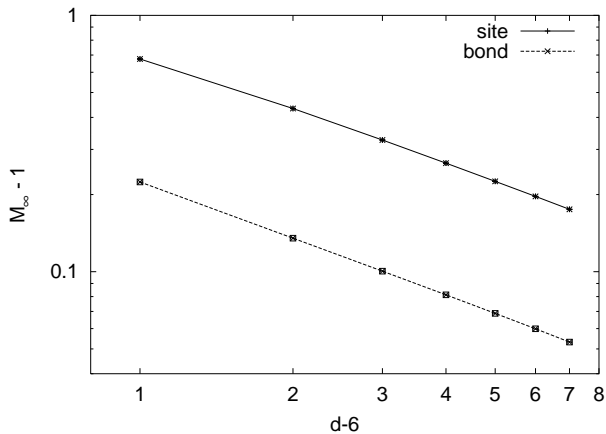,width=5.9cm,angle=270}
\caption{Log-log plot of $M_\infty-1$ against $d-d_c$. Statistical errors are 
smaller than the data symbols.}
\label{fig4.ps}
\end{figure}

The constants $M_\infty$ defined in Eq.(\ref{mt}) are plotted in Fig.4 against 
$d-6$ on doubly logarithmic scale. They seem to fall on parallel straight lines, 
suggesting a universal law $M_\infty -1 \sim (d-6)^{-a}$ with $a = 0.73 \pm 0.03$. 
But a closer look reveals that deviations from this are significant
(although they are small), suggesting that it holds neither for $d\to\infty$ nor 
for $d\to 6$ exactly.

\begin{figure}
\psfig{file=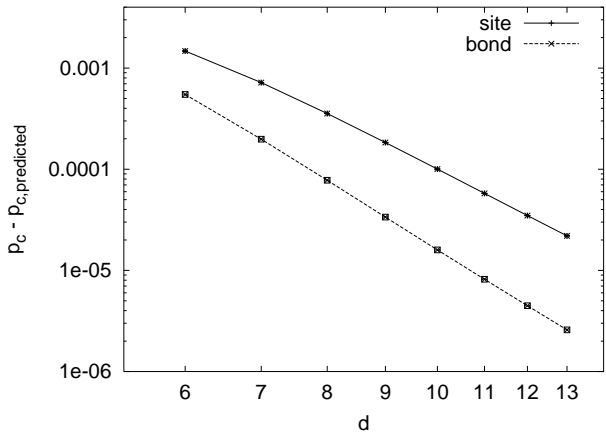,width=6.0cm,angle=270}
\caption{Log-log plot of the discrepancies between the simulation results and 
  Eqs.(\ref{seb}) and (\ref{ses}). Error bars are smaller than the sizes of the 
  symbols.}
\label{fig5.ps}
\end{figure}

Let us finally discuss the $p_c$ values given in Table 1. They should be compared 
to the predictions \cite{gr78} 
\begin{equation}
   p_{c,{\rm bond}} = s+5s^3/2+15s^4/2+57s^5 + \ldots     \label{seb}
\end{equation}
and \cite{gsr76}
\begin{equation}
   p_{c,{\rm site}} = s+3s^2/2+15s^3/4+83s^4/4 + \ldots     \label{ses}
\end{equation}
with $s = 1/(2d-1)$. The dots in these equations stand for higher powers of $s$. 
It was suggested in \cite{gsr76} that they can be approximated, for site 
percolation at least, by adding 2/3 of the last term,
\pagebreak
\begin{equation}
   p_{c,{\rm site}} \approx s+3s^2/2+15s^3/4+415s^4/12    \label{sesa}.
\end{equation}
The full series are presumably only asymptotic. It is thus a priori not clear 
whether any of these equations should be good approximations to the present 
data. From Table 1 we see that Eq.(\ref{sesa}) is excellent in the range 
studied here, but it has wrong asymptotic behaviour and should be worse 
than Eq.(\ref{ses}) for $d \ge 15$. As seen from Fig.5, the agreement with 
Eqs.(\ref{seb}) and (\ref{ses}) is indeed better than could have been expected:
For bond percolation the difference decreases roughly as $s^{7.1}$ (instead of 
$s^6$), while for site percolation it decreases as $s^{5.7}$ instead of $s^5$. 
Obviously the next terms in Eqs.(\ref{seb}) and (\ref{ses}) would be needed 
for a more detailed comparison.

Finally we should remind of several heuristic formulas for $p_c$ values on various 
lattices. All early ansatzes of this type were already refuted in \cite{gb84} 
because they contradicted Eqs.(\ref{seb}) or (\ref{ses}). More recently, such 
heuristics have been discussed again in \cite{gm01} and in the papers quoted there.
We have not attempted any detailed comparison in view of their complete lack of 
theoretical basis.

In summary, we have presented vastly improved estimates for percolation thresholds 
on high-dimensional hypercubic lattices. They should be compared to improved 
series expansions and/or rigorous bounds. At present such results are not available,
partly because it had seemed that they could not be compared to any numerical 
estimates. Apart from this, the methods used in the present paper should also 
be of use in other similar problems. These include simulations of percolation 
backbones, conductivity exponents, percolation on more exotic lattices, directed
percolation in high dimensions, and self avoiding walks. In all these cases both
the hashing and the variance reduction should be of help in simulating larger 
systems with higher precision.

I thank Walter Nadler and Hsiao-Ping Hsu for discussions and for carefully reading 
the manuscript.


\begin{thebibliography}{99}
\bibitem{sa94} D. Stauffer and A. Aharony, {\it Introduction to Percolation Theory},
   (Taylor \& Francis, London 1994).
\bibitem{cardy} J. Cardy, preprint math-ph/0103018 (2001).
\bibitem{s90} R. Sedgewick, {\it Algorithms in C} (Addison-Wesley, Reading 1990)
\bibitem{pzs01} G. Paul, R.M. Ziff, and H.E. Stanley, Phys. Rev. E {\bf 64}, 026115 (2001).
\bibitem{ziff} R.M. Ziff, Computers in Physics {\bf 12}, 385 (1998).
\bibitem{press} W.H. Press, B.P. Flannery, S.A. Teukolsky, and W.T. Vetterling, 
   {\it Numerical Recipes} (Cambridge Univ. Press, Cambridge 1997).
\bibitem{gr78} D.S. Gaunt and H. Ruskin, J. Phys. A {\bf 11}, 1369 (1978).
\bibitem{amah90} J. Adler, Y. Meir, A. Aharony, and A.B. Harris, Phys. Rev. B{\bf 41}, 9183 (1990).
\bibitem{m98} S. van der Marck, J. Phys. A {\bf 31}, 3449 (1998).
\bibitem{sz00} D. Stauffer and R.M. Ziff, Int. J. Modern Phys. C {\bf 11}, 205 (2000).
\bibitem{dickman} R. Dickman, Phys. Rev. E {\bf 60}, R2441 (1999).
\bibitem{balle} H.G. Ballesteros {\it et al.}, Phys. Lett. B {\bf 400}, 346 (1997).
\bibitem{footnote} In a branching process the evolution of 
   each particle and its offspring is independent of the past and of all other particles.
\bibitem{egg78} J.W. Essam, D.S. Gaunt, and A.J. Guttmann, J. Phys. A {\bf 11}, 1983 (1978).
\bibitem{g97} P. Grassberger, Phys. Rev. E{\bf 56}, 3682 (1997).
\bibitem{ghs94} P. Grassberger, R. Hegger, and L. Sch\"afer, J. Phys. A {\bf 27}, 7265 (1994).
\bibitem{po01} A.L. Owczarek and T. Prellberg, J. Phys. A {\bf 34}, 5773 (2001).
\bibitem{aah84} J. Adler, A. Aharony, and A.B. Harris, Phys. Rev. B{\bf 30}, 2832 (1984).
\bibitem{gsr76} D.S. Gaunt, M.F. Sykes, and H. Ruskin, J. Phys. A {\bf 9}, 1899 (1976).
\bibitem{gb84} D.S. Gaunt and R. Brak, J. Phys. A {\bf 17}, 1761 (1984).
\bibitem{gm01} S. Galam and A. Mauger, Phys. Rev. E {\bf 53}, 2177 (1996).
\end{thebibliography}
\end{document}